\documentclass[prl,reprint,twocolumn,showpacs,superscriptaddress]{revtex4-1}

\usepackage{bm,mathrsfs}
\usepackage{graphicx}
\usepackage{epsfig}
\usepackage{amsmath,bbm}
\usepackage{amsfonts,amssymb}
\usepackage{times}
\usepackage{dsfont}
\usepackage{enumitem}  
\usepackage{comment}
\usepackage[colorlinks,linkcolor=blue,citecolor=blue]{hyperref}
\newcommand{\eqn}[1]{
\begin{eqnarray}
	#1
\end{eqnarray}
}

\begin{document}
\title{Cavity Quantum Electrodynamics with Hyperbolic van der Waals Materials}

\author{Yuto Ashida}
\email{ashida@phys.s.u-tokyo.ac.jp}
\affiliation{Department of Physics, University of Tokyo, 7-3-1 Hongo, Bunkyo-ku, Tokyo 113-0033, Japan}
\affiliation{Institute for Physics of Intelligence, University of Tokyo, 7-3-1 Hongo, Tokyo 113-0033, Japan}
\author{Ata\c c $\dot{\mathrm{I}}$mamo$\breve{\mathrm{g}}$lu}
\affiliation{Institute of Quantum Electronics, ETH Zurich, CH-8093 Z{\"u}rich, Switzerland}
\author{Eugene Demler}
\affiliation{Institute for Theoretical Physics, ETH Zurich, 8093 Z{\"u}rich, Switzerland}

\begin{abstract}
The ground-state properties and excitation energies of a quantum emitter can be modified in the ultrastrong coupling regime of cavity quantum electrodynamics (QED) where the light-matter interaction strength becomes comparable to the cavity resonance frequency. Recent studies have started to explore the possibility of controlling an electronic material by embedding it in a cavity that confines electromagnetic fields in deep subwavelength scales. 
Currently, there is a strong interest in realizing ultrastrong-coupling cavity QED in the terahertz (THz)  part of the spectrum, since most of the elementary excitations of quantum materials are in this frequency range.
We propose and discuss a promising platform to achieve this goal based on a two-dimensional electronic material encapsulated by a planar cavity consisting of ultrathin polar van der Waals crystals. As a concrete setup, we show that nanometer-thick hexagonal boron nitride layers should allow one to reach the ultrastrong coupling regime for single-electron cyclotron resonance in a bilayer graphene. The proposed cavity platform can be realized by a wide variety of thin dielectric materials with hyperbolic dispersions. Consequently, van der Waals heterostructures hold the promise of becoming a versatile playground for exploring the ultrastrong-coupling physics of cavity QED materials.
\end{abstract}

\maketitle
Strong coupling regime of cavity quantum electrodynamics (QED), where the emitter-cavity coupling strength exceeds decay rates, has played a central role in quantum information science. For instance, cavity-mediated interaction between qubits has allowed for implementing two-qubit gates with high fidelity \cite{DL09,blais_circuit_2021}. Moreover, cavity-mediated Raman transitions have the potential to realize all-to-all coupling with tunable range and strength, providing an indispensable tool for quantum simulators \cite{SA12,VVD18}.  Recent studies have started to explore the possibility of further increasing the coupling strength so that it becomes comparable to elementary excitation energies. Many of the common simplifications in cavity QED fail in this regime, rendering theoretical analysis challenging \cite{HD10,DB182,LB20,ashida_cavity_2021}. A remarkable feature  is that ultrastrong coupling can alter the ground-state electronic properties due to virtual processes where both emitters and cavity photons are excited \cite{garcia-vidal_manipulating_2021,schlawin_cavity_2022,bloch_strongly_2022}. 
Consequently, a natural question to address is if and when the ultrastrong coupling regime can be attained and used to control material properties simply by cavity confinement in the absence of external driving.

A back-of-the-envelope estimate suggests that the ultrastrong coupling regime is out of reach in cavities supporting purely photonic excitations due to the smallness of the fine structure constant \cite{DMH07}.  However, this limit can be overcome in structured electromagnetic environments because hybridization with matter excitations enables one to control cavity frequencies independently of  wavelengths. For instance, in superconducting circuits, a large kinetic inductance allows for high-impedance electromagnetic excitations, leading to the single-electron ultrastrong coupling in the microwave range \cite{FDP17,YF172,forn-diaz_ultrastrong_2019,frisk_kockum_ultrastrong_2019}. 
 
Meanwhile, many of the elementary excitations in quantum materials are in the terahertz (THz) regime. Recent experimental and theoretical studies have shown the potential of utilizing cavity confinement as a means to control the phases of matter and chemical reactivity \cite{garcia-vidal_manipulating_2021,schlawin_cavity_2022,bloch_strongly_2022}. Specifically, earlier work demonstrated that cavity confinement modifies elementary excitations of many-body systems \cite{scalari_superconducting_2014,smolka_cavity_2014,maissen_ultrastrong_2014,zhang_collective_2016,BA17,keller_landau_2020,thomas_large_2021,RS22,KK22,FJ172,rokaj_quantum_2019,juraschek_cavity_2021,roman-roche_effective_2022,KS22,CJB22,orgiu_conductivity_2015,mann_manipulating_2018,MNS20,halbhuber_non-adiabatic_2020,hagenmuller_cavity-enhanced_2017,bartolo_vacuum-dressed_2018,li_electromagnetic_2020} and is expected to influence superconductivity \cite{schlawin_cavity-mediated_2019,curtis_cavity_2019,sentef_cavity_2018,gao_higgs_2021}, ferroelectricity \cite{ashida_quantum_2020,latini_ferroelectric_2021,LK22}, magnetotransport \cite{PB19,appugliese_breakdown_2022,rokaj_polaritonic_2022}, frustrated magnetism \cite{MK23,BEV22}, and topological properties \cite{wang_cavity_2019,MK22}. In view of the prospects of observing these intriguing phenomena in cavity QED materials, it is highly desirable to examine the feasibility of attaining the single-electron ultrastrong coupling at THz frequencies. This contrasts to the previous discussions of ultrastrong cavity QED that focused on multi-emitter systems, in which the coupling is enhanced by the factor $\sqrt{N}$ with $N$ being the number of emitters. 

\begin{figure}[b]
\includegraphics[width=86mm]{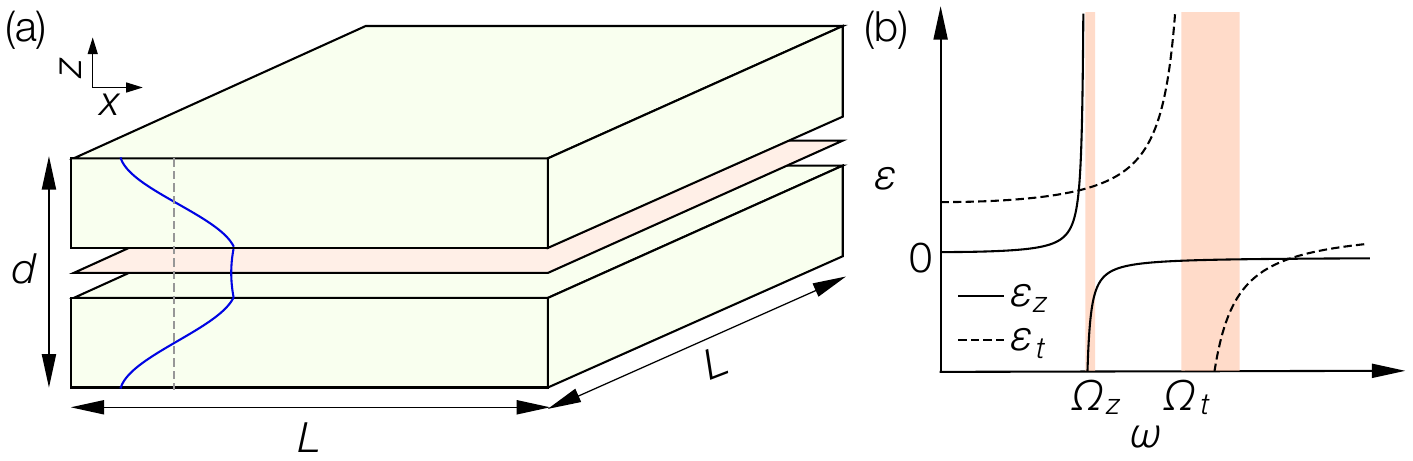} 
\caption{\label{fig_schematic}
Schematic figure illustrating the proposed planar cavity setup consisting of thin polar van der Waals crystals (green shaded) whose optical axis is along $z$ direction.  In the narrow air gap between the two slabs, a 2D material (red shaded) is inserted and the electron there can ultrastrongly couple to  electromagnetic fields of tightly confined hyperbolic polaritons (blue solid curve). The thickness (lateral extension) of surrounding materials is $d$ ($L$).  (b) Out-of-plane (solid curve) and in-plane (dashed curve) permittivities $\epsilon_{z,t}$ of hyperbolic materials. The Restrahlen bands (red shaded) appear above each of the out-of-plane and in-plane phonon resonances $\Omega_{z,t}$.}
\end{figure}

The aim of this Letter is to propose and analyze a planar cavity consisting of polar van der Waals (vdW) crystals as a promising platform for exploring the ultrastrong-coupling physics of cavity QED materials (Fig.~\ref{fig_schematic}a). We point out that one can attain a single-electron ultrastrong coupling by embedding a 2D material with electronic transitions in the THz regime into ultrathin hexagonal boron nitride ($\textit{h}$-BN) layers. 
A special feature here is that electrons in the 2D material couple to the electric field component of tightly confined hyperbolic phonon polaritons. The strong photon-phonon hybridization together with the low frequencies of polaritons then results in the significantly enhanced coupling strength over a broad range of momenta. This is challenging to achieve in conventional polar dielectrics with an isotropic dispersion, where sizable hybridization takes place in a limited range of momenta since light and matter are almost decoupled at large momenta due to the fast speed of light.

The proposed setup should be contrasted to existing approaches in several crucial aspects. In metallic platforms such as nanoplasmonic cavities \cite{MNS20}, inevitable Ohmic losses severely limit quality factors in deep subwavelength scales, and achieving the (ultra)strong coupling regime requires the collective enhancement, i.e., the single-electron ultrastrong coupling remains out of reach. Metallic structures also lead to static screening that could itself affect the electronic ground state. In superconducting circuits, the single-electron ultrastrong coupling has been attained, but the need of a large kinetic inductance restricts it to the microwave region which is much lower than THz frequencies relevant to excitations in real materials. In contrast, the present platform supports the confined hyperbolic polaritons in the THz regime while behaving as simple dielectrics at low frequencies, thus allowing for the THz single-electron ultrastrong coupling in the absence of Ohmic losses and strong screening. 

As a proof-of-concept demonstration, we will apply the proposed concept to a 2D electron gas with parabolic dispersion in the presence of the static magnetic field. We analyze the hybridization between hyperbolic polaritons and the cyclotron motion of the parabolic electrons. We show that the single-electron ultrastrong coupling regime is within the reach provided that the thicknesses of $\textit{h}$-BN layers are chosen to be nanometer-scale. This consideration is motivated by recent advances demonstrating ultrasmall mode volumes of hyperbolic phonon polaritons in $\textit{h}$-BN nanostructures \cite{CJD14,SD14,GA18,SD19,MEY22,SHH22}. While we present quantitative estimates for $\textit{h}$-BN nanocavities for the sake of concreteness, the proposed cavity scheme is applicable to the majority of polar vdW crystals \cite{GAK13}, which exhibit hyperbolic polaritons originating from distinct in- and out-of-plane infrared-active phonons.

{\it Hyperbolic phonon polaritons.---} 
We begin our analysis by reviewing the properties of a planar cavity made out of layered thin polar vdW materials. Due to the weakness of interlayer coupling, such materials naturally possess  two types of optical phonons corresponding to in-plane and out-of-plane ionic oscillations in the THz or mid-infrared regimes. This leads to the uniaxial anisotropy characterized by the out-of-plane (in-plane) dielectric permittivities $\epsilon_z$ ($\epsilon_t$). In the frequency windows above each of the phonon resonances, the two dielectric responses can have opposite signs (Fig.~\ref{fig_schematic}b). This unique feature leads to the hyperbolic isofrequency surfaces defined by the dispersion relation of the transverse magnetic modes \cite{JZ14,LP15},
\eqn{
\frac{|{\boldsymbol q}|^{2}}{\epsilon_{z}}+\frac{|{\boldsymbol \kappa}|^{2}}{\epsilon_{t}}=\frac{\omega^{2}}{\epsilon_{0}c^{2}},
} 
where $\boldsymbol q$ ($\boldsymbol \kappa$) is the in-plane (out-of-plane) wavevector, $\epsilon_0$ is the vacuum permittivity, and $c$ is the speed of light. The opposite signs of $\epsilon_{z,t}$ allow for excitations to have low frequencies even at large momenta. 
The resulting hybridized excitations are known as hyperbolic phonon polaritons. The corresponding dispersions are called the Reststrahlen bands of either Type I when $\epsilon_{z}<0,\epsilon_t>0$ or Type II when $\epsilon_{z}>0,\epsilon_t<0$.

\begin{figure}[b]
\includegraphics[width=86mm]{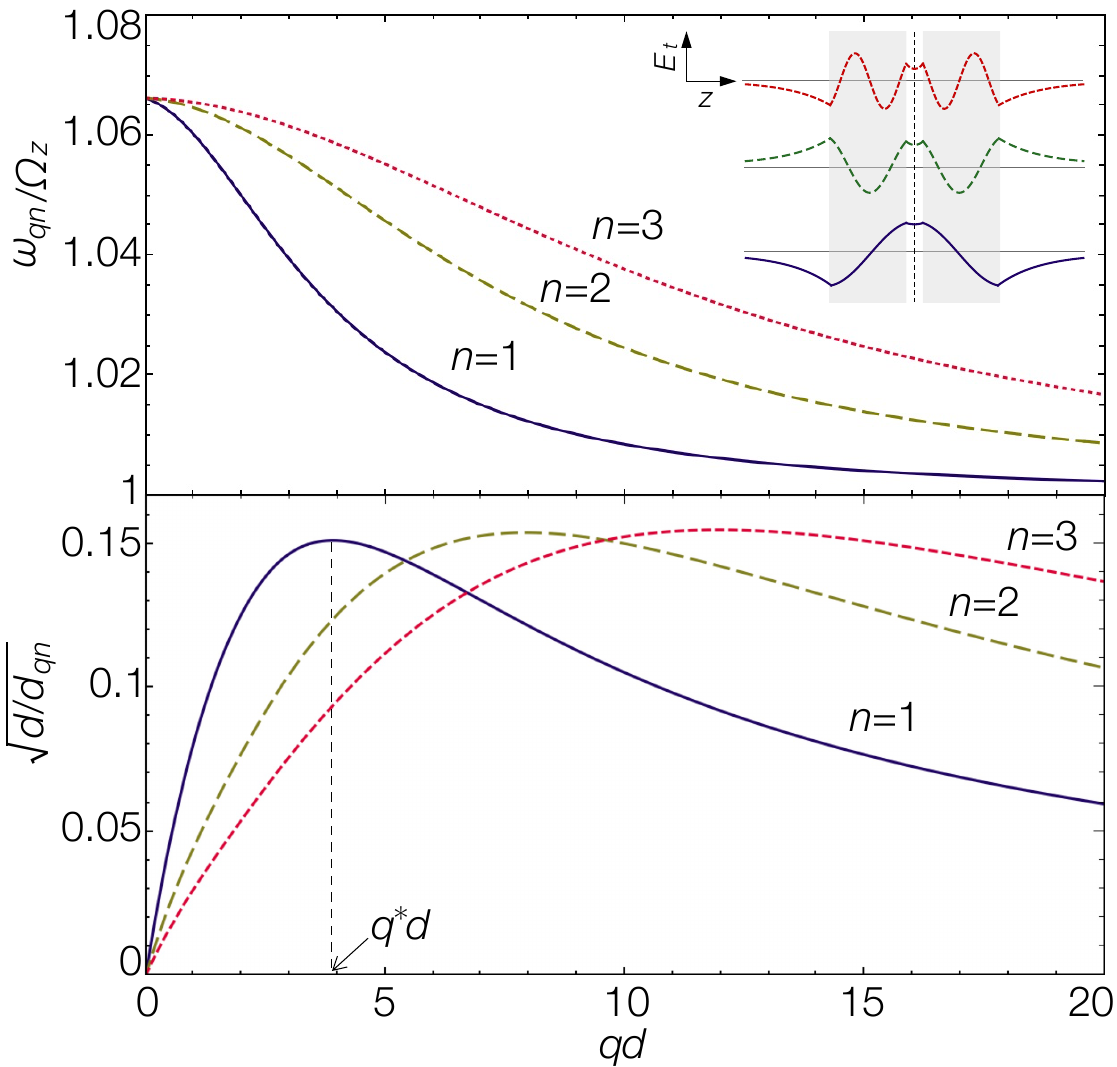} 
\caption{\label{fig_disp}
(Top) Dispersions $\omega_{{\boldsymbol q}n}$ of the hyperbolic phonon polaritons  in the planar cavity setting. The inset shows the spatial profiles of the in-plane electric fields along $z$ direction for each mode at $qd=4$. For the sake of visibility, only the three lowest modes are plotted. (Bottom) Inverse of the square root of the effective dimensionless confinement length, $\sqrt{d/d_{{\boldsymbol q}n}}$, characterizing the dimensionless single-electron coupling strength for each mode. The maximum value in the principal branch $n=1$ is reached at $q=q^*$.
}
\end{figure}

As a representative hyperbolic material, $\textit{h}$-BN has strong crystalline anisotropy leading to two spectrally distinct Reststrahlen bands  \cite{CJD14,SD14,GA18,SD19,MEY22,SHH22}. Below we focus on the electromagnetic couplings with Type I $\textit{h}$-BN hyperbolic modes $\omega_{\boldsymbol{q}n}$ lying in the narrow frequency window above the out-of-plane phonon frequency $\Omega_z=41.1\,{\rm THz}$ (cf. top panel in  Fig.~\ref{fig_disp}). As we discuss below, these modes exhibit several advantages for the purpose of attaining the ultrastrong couplings, such as their relatively low frequencies, sizable photon components over a broad range of momenta $q$, and ultraslow loss in deep subwavelength scales.

{\it Single-electron ultrastrong coupling with hyperbolic materials.---} 
 We consider the planar cavity setup where a 2D material (e.g., a bilayer graphene) is inserted into the narrow air gap between two thin $\textit{h}$-BN slabs whose thickness and lateral extensions are denoted by $d$ and $L$, respectively (Fig.~\ref{fig_schematic}a).  Our starting point is the following cavity QED Hamiltonian of the 2D parabolic electron interacting with hyperbolic phonon polaritons:
\eqn{
\hat{H}=\frac{\left[\hat{\boldsymbol{p}}+e\boldsymbol{A}_{s}(\hat{\boldsymbol{r}})+e\hat{\boldsymbol{A}}(\hat{\boldsymbol{r}})\right]^{2}}{2m}+\hat{H}_{\rm pol},
}
which can be derived from an effective theory of uniaxial polar dielectrics coupled to quantized electromagnetic fields in the Coulomb gauge  \cite{cohen-tannoudji_photons_1989,SM1}. 
Here, $e$ is the elementary charge, $m$ is the electron mass, $\hat{\boldsymbol{r}}$ and $\hat{\boldsymbol{p}}$ are the electron position and momentum operators in the 2D lateral directions, respectively, $\boldsymbol{A}_{s}$ represents an arbitrary static field, and $\hat{H}_{\rm pol}$ is the free polariton Hamiltonian,
\eqn{
\hat{H}_{{\rm pol}}=\sum_{\boldsymbol{q}n}\hbar\omega_{\boldsymbol{q}n}\hat{\gamma}_{\boldsymbol{q}n}^{\dagger}\hat{\gamma}_{\boldsymbol{q}n},
}
where $\hat{\gamma}_{\boldsymbol{q}n}$ ($\hat{\gamma}_{\boldsymbol{q}n}^\dagger$) annihilates (creates) a hyperbolic polariton with the in-plane wavevector $\boldsymbol q$ in the branch $n\in{\mathbb N}$.
The 2D vector field $\hat{\boldsymbol{A}}({\boldsymbol r})$ is obtained by projecting the vector potential onto the 2D plane where the electron is placed, and  can be expanded in terms of the polariton operators by
\eqn{\label{vecpo}
\hat{\boldsymbol{A}}(\hat{\boldsymbol{r}})=\sum_{\boldsymbol{q}n}{\cal A}_{\boldsymbol{q}n}\boldsymbol{e}_{\boldsymbol{q}}\left(\hat{\gamma}_{\boldsymbol{q}n}e^{i\boldsymbol{q}\cdot\hat{\boldsymbol{r}}}+\hat{\gamma}_{\boldsymbol{q}n}^{\dagger}e^{-i\boldsymbol{q}\cdot\hat{\boldsymbol{r}}}\right),
}
where ${\cal A}_{\boldsymbol{q}n}\simeq\sqrt{{\hbar}/({2L^{2}\epsilon_{0}\omega_{\boldsymbol{q}n}d_{\boldsymbol{q}n}})}$ is the  amplitude with the effective confinement length $d_{\boldsymbol{q}n}$ whose value is characterized by the polariton mode function. The effective polarization vector becomes $\boldsymbol{e}_{\boldsymbol{q}}\equiv\boldsymbol{q}/|{\boldsymbol q}|$ because, for the symmetric arrangement we assumed, the electric fields of the polaritons only have in-plane components along the propagation direction.
While the vector potential is originally a 3D transverse vector field  in the Coulomb gauge, it can effectively acquire longitudinal components when projected onto the 2D tangential plane. 

Figure~\ref{fig_disp} shows the results for Type I $\textit{h}$-BN hyperbolic modes. The dispersions (top panel) start from the longitudinal phonon frequency and saturate to  $\Omega_z$ at high $q\equiv|{\boldsymbol q}|$. Naturally, these hybridized modes are almost purely longitudinal (transverse) phonon excitations in the limit $q\to 0$ ($q\to\infty$), which do not allow for a strong coupling to electrons in the air gap. Crucially, however, except for these two limits, there still exist the nonvanishing photon contributions leading to the sizable electric-dipole couplings at intermediate momenta (bottom panel). This is because the in-plane component, which couples to the 2D electron, has almost equal photonic  and phononic contents while the out-of-plane component is predominantly phonon-like \cite{SM1}.  

\begin{figure*}[t]
\includegraphics[width=175mm]{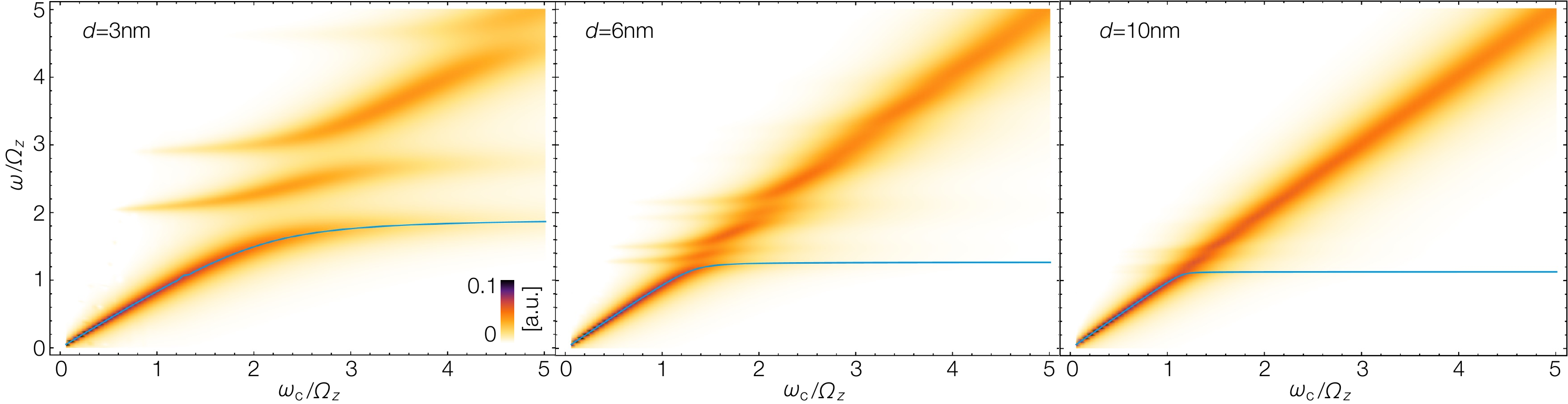} 
\caption{\label{fig_abs}
Magnetoabsorption spectra $A(\omega)$ of the 2D electron in the cavity plotted against a cyclotron resonance $\omega_c=eB/m$ at different cavity thicknesses $d$. The blue solid curve in each panel corresponds to the bound-state energy of the Landau polariton. We impose the periodic boundary conditions on the lateral directions and fix the aspect ratio $L/d=\pi$. We use the $\textit{h}$-BN parameters in Ref.~\cite{CJD14} where the out-of-plane and in-plane phonon frequencies are $\Omega_{z}=23.3\;{\rm THz}$ and $\Omega_{t}=41.1\;{\rm THz}$, the corresponding oscillator strengths are $\eta_{z}=0.37\Omega_{z}$ and $\eta_{t}=0.61\Omega_{t}$, and the  infinite-frequency permittivities are $\epsilon_{z\infty}=2.95$ and $\epsilon_{t\infty}=4.87$ \cite{SM1}. We set the in-plane momentum cutoff $\Lambda=2{\rm \,nm}^{-1}$. 
}
\end{figure*}

To further proceed, we use the unitary transformation $\hat{U}=e^{-i\hat{\boldsymbol{r}}\cdot\hat{\boldsymbol{P}}_{b}/\hbar}$ with $\hat{\boldsymbol{P}}_{b}=\sum_{\boldsymbol{q}n}\hbar\boldsymbol{q}\hat{\gamma}_{\boldsymbol{q}n}^{\dagger}\hat{\gamma}_{\boldsymbol{q}n}$ \cite{LLP53}, resulting in the Hamiltonian $\hat{H}_{U}\equiv\hat{U}^{\dagger}\hat{H}\hat{U}$ given by
\eqn{\label{HU}
\hat{H}_{U}=\frac{\left[\hat{\boldsymbol{p}}\!+\!e\boldsymbol{A}_{s}(\hat{\boldsymbol{r}})\!+\!e\hat{\boldsymbol{A}}(\boldsymbol{0})\!-\!\hat{\boldsymbol{P}}_{b}\right]^{2}}{2m}+\hat{H}_{\rm pol}.
}
In this way, the electron-coordinate dependence in the quantized vector potential can be eliminated at the expense of generating the polariton momentum $\hat{\boldsymbol{P}}_{b}$, leading to the nonlocal interaction among polaritons mediated by the electron. The dimensionful coupling strength between the electron and each of the dynamical quantized electromagnetic modes  is given by 
\eqn{
g_{\boldsymbol{q}n}&=&e{\cal A}_{\boldsymbol{q}n}\sqrt{\frac{\omega_{\boldsymbol{q}n}}{m\hbar}}.
}
Since $g_{\boldsymbol{q}n}$ characterizes the single-electron coupling strength rather than the collective one, it depends on the lateral size $L$ through  $g_{\boldsymbol{q}n}\propto L^{-1}$. Consequently,  a natural measure for the effective coupling strength between the 2D electron and the continuum of polariton modes is given by the integrated value $g_{{\rm eff}}\equiv\sqrt{\sum_{\boldsymbol{q}n}g_{\boldsymbol{q}n}^{2}}$, which scales as $g_{\rm eff}\propto {\rm O}(L^0)$ \cite{ashida_nonperturbative_2022}. 

Previously, the deep strong coupling regime has been experimentally realized in superconducting circuits, where $g_{\rm eff}$ becomes comparable to the microwave photon frequency. In the present setting, the use of ultrathin $\textit{h}$-BN slabs with nanometer-scale thicknesses enables one to reach those regimes in the THz range, where $g_{{\rm eff}}$ becomes comparable to or even exceeds $\omega_{\boldsymbol{q}n}$.  For instance, using the $\textit{h}$-BN parameters \cite{CJD14}, we estimate coupling strengths  of order $g_{\boldsymbol{q}n}/\Omega_{z}=1.7\times\sqrt{(10{\rm \,nm})^{3}/(L^{2}d_{\boldsymbol{q}n})}$, which, together with the results of the effective confinement length $d_{{\boldsymbol q}n}$ in Fig.~\ref{fig_disp}b, can lead to $g_{\rm eff}\sim\Omega_z$ in nanoscale heterostructures. More specifically, one can attain $g_{\rm eff}\simeq 2.0\,\Omega_z$ for the $n=1$ principal branch  when the cavity thickness (in-plane momentum cutoff) is set to be $d=5{\rm \,nm}$ ($\Lambda=2{\rm \,nm}^{-1}$). As demonstrated below, one important consequence of attaining $g_{\rm eff}\sim\Omega_z$ is that the electron mixes the otherwise independent cavity modes and creates a localized state at a frequency below the cavity resonance. We emphasize that this key feature is largely insensitive to a choice of the lateral size $L$ and thus remains even in the 2D thermodynamic limit $L/d\to\infty$.

{\it Application to a 2D electron under the magnetic field.---} 
As a proof-of-concept demonstration, we now focus on a prototypical setting of ultrastrong-coupling physics, namely, a 2D parabolic electron subject to a static perpendicular magnetic field. From now on, we consider the $n=1$ principal hyperbolic mode  that  most strongly couples to the cyclotron motion of the electron, and abbreviate the subscript $n$ for the sake of notational simplicity. We choose the symmetric gauge, $\boldsymbol{A}_{s}(\boldsymbol{r})=(-By/2,Bx/2)^{{\rm T}}$, with the magnetic field $B$ and  introduce the annihilation operator of Landau levels by
\eqn{
\hat{a}=\frac{1}{\sqrt{2}}\left[\frac{l_{B}}{\hbar}\left(\hat{p}_{x}-i\hat{p}_{y}\right)-\frac{i}{2l_{B}}\left(\hat{x}-i\hat{y}\right)\right],
}
where $l_{B}=\sqrt{\hbar/(eB)}$ is the magnetic length. 
Using the cyclotron frequency $\omega_{c}=eB/m$ and the operator $\hat{\boldsymbol{\pi}}=\frac{1}{\sqrt{2}}\left(\hat{a}+\hat{a}^{\dagger},i(\hat{a}-\hat{a}^{\dagger})\right)^{{\rm T}}$, we obtain
\eqn{\label{HU2}
\hat{H}_{U}&\!=\!&\frac{\hbar\omega_{c}}{2}\left(\hat{\boldsymbol{\pi}}\!+\!\sum_{\boldsymbol{q}}\boldsymbol{e}_{\boldsymbol{q}}\left[c_{\boldsymbol{q}}\left(\hat{\gamma}_{\boldsymbol{q}}\!+\!\hat{\gamma}_{\boldsymbol{q}}^{\dagger}\right)\!-\!ql_{B}\hat{\gamma}_{\boldsymbol{q}}^{\dagger}\hat{\gamma}_{\boldsymbol{q}}\right]\right)^{2}\!\!+\!\hat{H}_{\rm pol},\nonumber\\}
where $c_{\boldsymbol{q}}=g_{\boldsymbol{q}}/\sqrt{\omega_{\boldsymbol{q}}\omega_{c}}$ is the dimensionless coefficient characterizing the coupling strength of the cyclotron motion to the hyperbolic polaritons with momentum $\boldsymbol q$. In the long-wavelength limit $ql_{B}\to0$, the polariton-polariton interaction in Eq.~\eqref{HU2} disappears and the problem reduces to the quadratic one. As demonstrated below, however, this interaction term can in general contribute to the dynamics and affect the absorption spectrum especially in the ultrastrong coupling regimes. This is because the electron couples most strongly with the electromagnetic modes at a finite momentum around $q\sim q^*$ (cf. Fig.~\ref{fig_disp}b) whose corresponding length scale $1/q^*$ is comparable to the magnetic length $l_B$ near the cyclotron resonance.

The low-energy excitations can be studied by analyzing the magnetoabsorption spectrum
\eqn{
A(\omega)={\rm Re}\left[\int_{0}^{\infty}e^{i\omega t}\langle{\rm GS}|\hat{a}e^{-i\hat{H}t}\hat{a}^{\dagger}|{\rm GS}\rangle\right],
}
where $|{\rm GS}\rangle$ is the ground state. To reveal its qualitative features, we perform a simple variational analysis as follows. Specifically, we first determine the variational ground state of Eq.~\eqref{HU2} in the form of a product of coherent states. We then expand the Hamiltonian around this state, obtain the fluctuations up to the quadratic terms, and determine the excitation spectrum via the exact diagonalization of the effective quadratic Hamiltonian \cite{SM1}. 

Figure~\ref{fig_abs} shows the obtained magnetoabsorption spectra, where the cavity thickness $d$ is varied while the aspect ratio $L/d$ is kept constant. The blue solid curve in each panel shows the bound-state energy of the Landau polariton. As the thickness is decreased, the spectrum starts to exhibit the anticrossings around the cyclotron resonance. In particular, when the cavity length becomes a few nanometers, there emerge the large separations between the branches that are comparable to the elementary excitation energies; this is a hallmark of the ultrastrong coupling regime. 

As discussed before, a key feature here is the formation of the dressed bound state consisting of the electron and the localized polaritons, which manifests itself as the anticrossed lower branch in the spectrum (cf. the blue curve in Fig.~\ref{fig_abs}). Importantly, this feature remains independently of lateral size $L$, while the effect of increasing  $L/d$ is the appearance of a continuum of cavity modes above the lower branch \cite{SM1}.
It is worthwhile to note that the positions of the anticrossings are shifted above the bare resonances $\omega_{c}/\Omega_z\simeq 1$. This upward shift originates from the renormalization of the effective polariton energies due to the repulsive polariton-polariton interaction. Since the latter comes from the spatial dependence of the vector potential, this effect is absent in the long-wavelength limit. We also remark that the appearance of the multiple anticrossed branches in Fig.~\ref{fig_abs} are due to the discretized in-plane momentum $q$, whose value is set by the periodic boundary conditions in the lateral directions.

{\it Discussions.---} 
The proposed platform for ultrastrong-coupling cavity QED materials can be realized by various polar vdW materials exhibiting hyperbolic phonon polaritons, including Bi$_{2}$Se$_3$, Bi$_{2}$Te$_{3}$, MoS$_{2}$ and MoO$_{3}$ as well as $\textit{h}$-BN \cite{EM14,GAK13,SJ14,MW18,ZZ18,ZZ19}. Thus, by confining materials in the  cavity consisting of these vdW structures,  one can strongly couple electronic excitations to quantized electromagnetic modes in a wide spectral range from mid- or far-infrared to THz regimes. Moreover, the layered nature of vdW crystals should readily allow one to tune the cavity coupling strengths by controlling the thickness of the surrounding crystals. 

While polaritons in these materials exhibit ultralow loss originating from multi-phonon or disorder-induced scatterings \cite{CJD14,SD14,GA18,SD19,MEY22,SHH22,MW18,ZZ18,ZZ19}, it merits further study to explore if and when one could engineer such dissipation to realize phases or dynamics unique to open systems \cite{YAreview,FM21}. 
Our analysis focused on 2D materials and it remains an interesting open problem how one can realize ultrastrong coupling between light and 3D bulk materials if at all possible.
Finally, in quantum information science, a material excitation consisting of localized polaritons might find applications in  photon storage, quantum memory, or nondissipative emitter interactions \cite{LA09,GTA15}.  

In summary, we showed that the planar cavity consisting of vdW heterostructures provides a promising platform to attain the single-electron ultrastrong coupling in the THz or mid-infrared regions. As a proof-of-concept demonstration, we presented the analysis of the magnetoabsorption spectrum for the 2D electron confined in the $\textit{h}$-BN cavity, where the coupling can be ultrastrong provided that the thicknesses are judiciously controlled. We expect that our results open a way to study a variety of recent predictions in the emerging field of cavity QED materials.

\begin{acknowledgments}
We are grateful to  Iliya Esin, Ilya Esterlis, Tim Kaxiras, Igor Khanonkin, Frank Koppens, Kanta Masuki, Gil Refael, Tao Shi, and Kenji Yasuda for fruitful discussions. Y.A. acknowledges support from the Japan Society for the Promotion of Science through Grant No.~JP19K23424 and from JST FOREST Program (Grant Number~JPMJFR222U, Japan). A.I. was supported by the Swiss National Science Foundation (SNSF) under Grant Number 200020\_207520. E.D. acknowledges support from the ARO grant ``Control of Many-Body States Using Strong Coherent Light-Matter Coupling in Terahertz Cavities" and the SNSF project 200021\_212899.
\end{acknowledgments}

\bibliography{reference}

\widetext
\pagebreak
\begin{center}
\textbf{\large Supplementary Materials}
\end{center}

\renewcommand{\theequation}{S\arabic{equation}}
\renewcommand{\thefigure}{S\arabic{figure}}
\renewcommand{\bibnumfmt}[1]{[S#1]}
\setcounter{equation}{0}
\setcounter{figure}{0}

\subsection{Derivation of the effective cavity QED Hamiltonian}

We here derive the effective Hamiltonian of the proposed cavity QED system discussed in the main text. To this end, we start from a general microscopic model of the 2D electron coupled to quantized electromagnetic fields in uniaxial polar dielectrics in the Coulomb gauge \cite{cohen-tannoudji_photons_1989,ashida_quantum_2020}:
\eqn{\hat{H}=\frac{\left[\hat{\boldsymbol{p}}+e\boldsymbol{A}_{s}(\hat{x},\hat{y})+e\hat{\boldsymbol{A}}(\hat{x},\hat{y},z=0)\right]^{2}}{2m}+\hat{H}_{0},\;\;\;\hat{\boldsymbol{p}}=\left(\begin{array}{c}
\hat{p}_{x}\\
\hat{p}_{y}
\end{array}\right),}
where $m$ is the electron mass, $e$ is the elementary charge, and $\boldsymbol{A}_{s}$ is an arbitrary static external field. We note that the 2D vector field $\hat{\boldsymbol{A}}(x,y,z=0)$ is defined by the projection of the following 3D quantized vector potential onto the 2D tangential plane at $z=0$:
\eqn{
\hat{\boldsymbol{A}}(\boldsymbol{r})=\sum_{\boldsymbol{k}\lambda}\sqrt{\frac{\hbar}{2V\epsilon_{0}ck}}\left[\hat{a}_{\boldsymbol{k}\lambda}\boldsymbol{\epsilon}_{\boldsymbol{k}\lambda}e^{i\boldsymbol{k}\cdot\boldsymbol{r}}+{\rm H.c.}\right],
}
where $V$ is the volume of the entire space, and $\hat{a}_{\boldsymbol{k}\lambda}$ ($\hat{a}_{\boldsymbol{k}\lambda}^\dagger$) is the annihilation (creation) operator of photons with wavevector $\boldsymbol{k}$ and polarization $\lambda$; note that these excitations correspond to bare photons defined in the absence of dielectric materials. In the Coulomb gauge, the orthonormal transverse polarization vectors $\boldsymbol{\epsilon}_{\boldsymbol{k}\lambda}$ satisfy  $\boldsymbol{k}\cdot\boldsymbol{\epsilon}_{\boldsymbol{k}\lambda}=0$ and $\boldsymbol{\epsilon}_{\boldsymbol{k}\lambda}^\dagger\boldsymbol{\epsilon}_{\boldsymbol{k}\nu}=\delta_{\lambda\nu}$. The term $\hat{H}_{0}$ is the quadratic Hamiltonian consisting of the energies of the quantized electromagnetic fields and the polar phonons in dielectrics: 
\eqn{
\hat{H}_{0}=\int d^{3}r\left[\frac{\hat{\boldsymbol{\Pi}}^{2}}{2\boldsymbol{\epsilon}(z)}+\frac{\epsilon_0 c^2}{2}\left(\nabla\times\hat{\boldsymbol{A}}\right)^{2}\right]+\int_{{\rm D}}\frac{d^{3}r}{v}\sum_{i\in\{x,y,z\}}\left[\frac{\left(\hat{\pi}_{i}-Z^{*}e\hat{A}_{i}\right)^{2}}{2M_{i}}+\frac{1}{2}M_{i}\Omega_{i}^{2}\hat{\phi}_{i}^{2}+\frac{\left(Z^{*}e\hat{\phi}_{i}^{\parallel}\right)^{2}}{2\epsilon_{i}(z)v}\right],
}
where $\hat{\boldsymbol{\Pi}}(\boldsymbol{r})$ is the canonically conjugate vector field of $\hat{\boldsymbol{A}}(\boldsymbol{r})$, $v$ is the unit-cell volume of the dielectrics, $\hat{\boldsymbol{\phi}}(\boldsymbol{r})$ is a vector field representing polar optical phonons, $\hat{\boldsymbol{\pi}}(\boldsymbol{r})$ is its canonically conjugate vector field, $Z^*e$ is the effective charge of polar phonons, and $\int_{{\rm D}}d^{3}r$ represents the integration over the dielectric regions, 
$\int_{{\rm D}}d^{3}r=\int d^{2}r\left(\int_{-d/2}^{-\delta/2}+\int_{\delta/2}^{d/2}\right)dz$, with $\delta$ being the thickness of the narrow air gap that is assumed to be much shorter than the cavity thickness $d$. 
We also define the permittivity tensor $\boldsymbol{\epsilon}(z)$ by
\eqn{
\boldsymbol{\epsilon}(z)={\rm diag}\left(\epsilon_{t\infty},\epsilon_{t\infty},\epsilon_{z\infty}\right)\vartheta\left(\frac{d}{2}-|z|\right)\vartheta\left(|z|-\frac{\delta}{2}\right)+\epsilon_{0}I\left[\vartheta\left(|z|-\frac{d}{2}\right)+\vartheta\left(\frac{\delta}{2}-|z|\right)\right],
}
where $\epsilon_{t(z)\infty}$ is the in-plane (out-of-plane) permittivity in the infinite-frequency limit and $I$ is the identity matrix. We denote the in-plane  masses and phonon frequencies by $M_{x,y}=M_{t}$ and $\Omega_{x,y}=\Omega_{t}$, respectively. We remark that the permittivity in the air gap region can in general take a value different from the vacuum value depending on a specific choice of the cavity geometry, while we take it to be $\epsilon_0$ in the present work for the sake of simplicity.

To derive the effective Hamiltonian, we first need to identify a polariton eigenmode that diagonalizes the quadratic part $\hat{H}_0$. This can be done by solving the macroscopic Maxwell's equations derived from the above model. We remark that the influence of the static external field can be neglected when constructing such polariton modes. Without loss of generality, we consider eigenmodes propagating along the $x$ direction with in-plane momentum $\boldsymbol{q}=(q,0,0)^{{\rm T}}$. Also, we recall that we are interested in the hyperbolic eigenmodes that strongly couple to the 2D electron moving on the $z=0$ plane. Such modes correspond to the P-polarized (or equivalently, the transverse magnetic) modes with the even parity for the tangential component, which satisfy the following conditions: $[\boldsymbol{B}]_{y}\neq0, [\boldsymbol{B}]_{x,z}=0$,  $[\boldsymbol{E}]_{y}=0, [\boldsymbol{E}]_{x,z}\neq0$, and $E_{x}(z)=E_{x}(-z), E_{z}(z)=-E_{z}(-z)$. The resulting eigenequations that characterize these modes are
\eqn{
\frac{q^{2}}{\epsilon_{z}(\omega)}+\frac{\kappa^{2}}{\epsilon_{t}(\omega)}&=&\frac{\omega^{2}}{\epsilon_{0}c^{2}},\label{eig1}\\
\tan\left(\frac{\kappa d}{2}\right)&=&-\frac{\kappa}{\epsilon_{t}\sqrt{q^{2}-\frac{\omega^{2}}{c^{2}}}},\label{eig2}
}
where $\kappa$ is the out-of-plane momentum and 
the in-plane and out-of-plane permittivities $\epsilon_{t,z}(\omega)$ are given by
\eqn{
\epsilon_{t}(\omega)=\epsilon_{t\infty}\left(1+\frac{\eta_{t}^{2}}{\Omega_{t}^{2}-\omega^{2}}\right),\;\;\epsilon_{z}(\omega)=\epsilon_{z\infty}\left(1+\frac{\eta_{z}^{2}}{\Omega_{z}^{2}-\omega^{2}}\right).
}
Here, we denote the coupling strengths of the dielectrics by  
\eqn{
\eta_{t}=\sqrt{\frac{\left(Z^{*}e\right)^{2}}{\epsilon_{t\infty}M_{t}v}},\;\;\;
\eta_{z}=\sqrt{\frac{\left(Z^{*}e\right)^{2}}{\epsilon_{z\infty}M_{z}v}}.
}
For the $\textit{h}$-BN crystals considered in the main text, we use the parameters in Ref.~\cite{CJD14} where the out-of-plane and in-plane phonon frequencies  are $\Omega_{z}=23.3\;{\rm THz}$ and $\Omega_{t}=41.1\;{\rm THz}$, the corresponding couplings are $\eta_{z}=0.37\Omega_{z}$ and $\eta_{t}=0.61\Omega_{t}$, and the out-of-plane and in-plane  infinite-frequency permittivities are $\epsilon_{z\infty}=2.95$ and $\epsilon_{t\infty}=4.87$, respectively.

\begin{figure}[t]
\includegraphics[width=170mm]{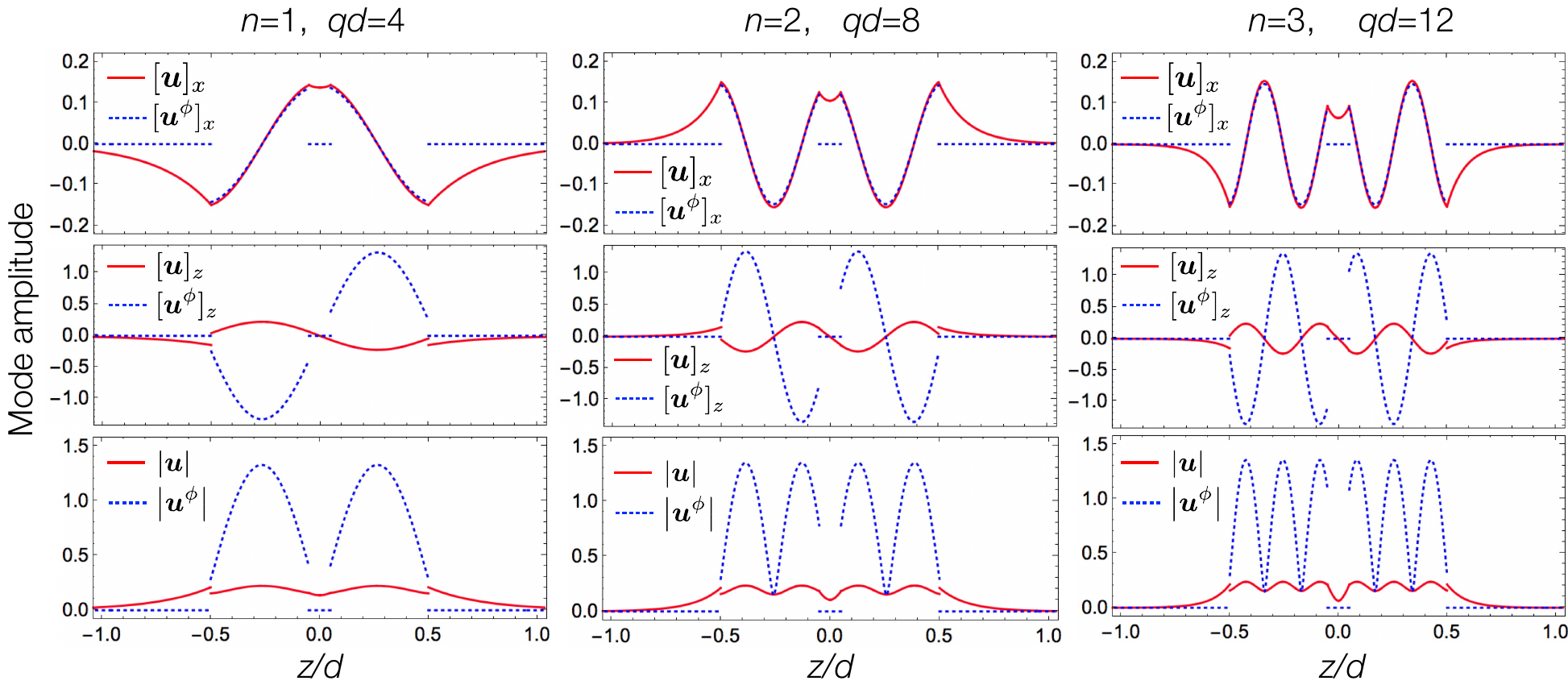} 
\caption{\label{fig_s1}
Mode profiles of the $\textit{h}$-BN Type I hyperbolic phonon polaritons. The top, middle, and bottom panels plot the spatial dependences of the mode functions along the thickness direction for their in-plane components, out-of-plane components, and root mean squares, respectively. The red solid (blue dotted) curves correspond to the electric-field (phonon) mode functions $\boldsymbol{u}_{\boldsymbol{q}n}$ ($\boldsymbol{u}_{\boldsymbol{q}n}^\phi$).  }
\end{figure}

Solving the eigenequations \eqref{eig1} and \eqref{eig2} with respect to $\kappa$ and $\omega$ for a given $q$, we obtain a hyperbolic dispersion  $\omega_{\boldsymbol{q}n}$ associated with the out-of-plane momentum $\kappa_{\boldsymbol{q}n}$, which are labeled by the discrete number $n$. The corresponding mode function, $\boldsymbol{u}_{\boldsymbol{q}n}$,  characterizes the spatial profile of the electric field and is given by
\eqn{
\boldsymbol{u}_{\boldsymbol{q}n}&=&f_{\boldsymbol{q}n}\left[\cos(\kappa_{\boldsymbol{q}n}z)\boldsymbol{e}_{x}-\frac{i\epsilon_{t}q}{\epsilon_{z}\kappa_{\boldsymbol{q}n}}\sin(\kappa_{\boldsymbol{q}n}z)\boldsymbol{e}_{z}\right]e^{iqx}\vartheta\left(\frac{d}{2}-|z|\right)\vartheta\left(|z|-\frac{\delta}{2}\right)\nonumber\\
&+&f'_{\boldsymbol{q}n}\sum_{s=\pm1}\left[\boldsymbol{e}_{x}+is(q/\nu_{\boldsymbol{q}n})\boldsymbol{e}_{z}\right]e^{iqx-s\nu_{\boldsymbol{q}n}z}\vartheta\left(|z|-\frac{d}{2}\right)\nonumber\\
&+&f''_{\boldsymbol{q}n}\left[\cosh(\nu_{\boldsymbol{q}n}z)\boldsymbol{e}_{x}-i(q/\nu_{\boldsymbol{q}n})\sinh(\nu_{\boldsymbol{q}n}z)\boldsymbol{e}_{z}\right]e^{iqx}\vartheta\left(\frac{\delta}{2}-|z|\right),
}
where $\nu_{\boldsymbol{q}n}=\sqrt{q^{2}-{\omega_{\boldsymbol{q}n}^{2}}/{c^{2}}}$ and  the amplitude coefficients satisfy the following conditions:
\eqn{
f_{\boldsymbol{q}n}/f'_{\boldsymbol{q}n}&=&e^{-\nu_{\boldsymbol{q}n}d/2}/\cos(\kappa_{\boldsymbol{q}n}d/2),\\ 
f_{\boldsymbol{q}n}/f''_{\boldsymbol{q}n}&=&\cosh\left(\nu_{\boldsymbol{q}n}\delta/2\right)/\cos(\kappa_{\boldsymbol{q}n}\delta/2).}
Without loss of generality, we choose $f_{\boldsymbol{q}n}>0$. 
The corresponding phonon mode function is given by
\eqn{
\boldsymbol{u}_{\boldsymbol{q}n}^{\phi}=f_{\boldsymbol{q}n}\left[\frac{\eta_{t}\Omega_{t}}{\Omega_{t}^{2}-\omega_{\boldsymbol{q}n}^{2}}\cos(\kappa_{\boldsymbol{q}n}z)\boldsymbol{e}_{x}-\frac{i\epsilon_{t}q}{\epsilon_{z}\kappa_{\boldsymbol{q}n}}\frac{\eta_{z}\Omega_{z}}{\Omega_{z}^{2}-\omega_{\boldsymbol{q}n}^{2}}\sin(\kappa_{\boldsymbol{q}n}z)\boldsymbol{e}_{z}\right]e^{iqx}\vartheta\left(\frac{d}{2}-|z|\right)\vartheta\left(|z|-\frac{\delta}{2}\right).
}
We then impose the normalization condition,
\eqn{
\int_{-\infty}^{\infty}dz\left[\left|\boldsymbol{u}_{\boldsymbol{q}n}(z)\right|^{2}+\left|\boldsymbol{u}_{\boldsymbol{q}n}^{\phi}(z)\right|^{2}\right]=1,
}
which fixes the amplitude coefficients. 

The results for the three lowest branches of the $\textit{h}$-BN Type I hyperbolic modes are shown in Fig.~\ref{fig_s1}. 
We remark that, on the $z=0$ plane where the 2D material is inserted, the electric fields have only the in-plane components along the propagation direction. 
We also note that, in the air gap region, the electric field is written as the curl of the magnetic field and thus it is purely transverse, i.e., $\boldsymbol{u}_{\boldsymbol{q}n}\propto\nabla\times\left(\sinh(\nu_{\boldsymbol{q}n}z)e^{iqx}\boldsymbol{e}_{y}\right)$ for $|z|<\frac{\delta}{2}$. Since we assume the narrow gap $\nu,\kappa\ll\delta^{-1}$, we can introduce the effective confinement length $d_{\boldsymbol{q}n}$ through the relation $|\boldsymbol{u}_{\boldsymbol{q}n}(z=0)|\simeq f_{\boldsymbol{q}n}\equiv 1/\sqrt{d_{\boldsymbol{q}n}}$, which characterizes the coupling strength $g_{\boldsymbol{q}n}$ of the 2D electron on the $z=0$ plane as discussed in the main text.

\subsection{Details about the mean-field analysis of the magnetoabsorption spectrum}

\begin{figure}[b]
\includegraphics[width=170mm]{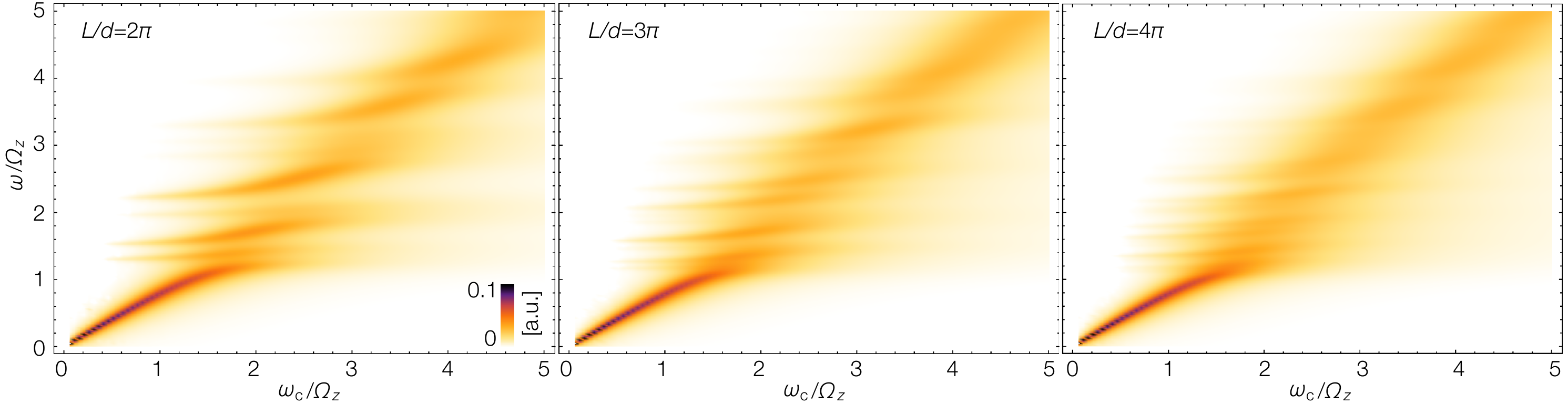} 
\caption{\label{fig_s2}
Magnetoabsorption spectra $A(\omega)$ at different aspect ratios $L/d$. The cavity thickness $d$ is fixed to be $d=3{\rm\,nm}$. The in-plane momentum $\boldsymbol{q}$ is discretized as in Eq.~\eqref{qdisc}. The results are plotted for the $\textit{h}$-BN parameters and the in-plane momentum cutoff $\Lambda=2{\,{\rm nm}^{-1}}$.
 }
\end{figure}

We here provide technical details about the mean-field variational analysis of the magnetoabsorption spectrum discussed in the main text. To this end, we first  perform an additional unitary transformation,
\eqn{\hat{V}=e^{\sum_{\boldsymbol{q}n}\frac{c_{\boldsymbol{q}n}}{l_{B}}\left(\hat{\gamma}_{\boldsymbol{q}n}^{\dagger}-\hat{\gamma}_{\boldsymbol{q}n}\right)},\;\;\hat{V}^{\dagger}\hat{\gamma}_{\boldsymbol{q}n}\hat{V}=\hat{\gamma}_{\boldsymbol{q}n}+\frac{c_{\boldsymbol{q}n}}{l_{B}},}
which displaces the polariton operators. Using the relation $\sum_{\boldsymbol{q}n}\boldsymbol{q}c_{\boldsymbol{q}n}^{2}=\boldsymbol{0}$, the Hamiltonian in Eq.~(8) in the main text is transformed to
\eqn{
\hat{H}_{VU}=\hat{V}^{\dagger}\hat{H}_{U}\hat{V}=\frac{\hbar\omega_{c}}{2}\left(\hat{\boldsymbol{\pi}}-\sum_{\boldsymbol{q}n}\boldsymbol{q}l_{B}\hat{\gamma}_{\boldsymbol{q}n}^{\dagger}\hat{\gamma}_{\boldsymbol{q}n}\right)^{2}+\sum_{\boldsymbol{q}n}\hbar\omega_{\boldsymbol{q}n}\left[\hat{\gamma}_{\boldsymbol{q}n}^{\dagger}\hat{\gamma}_{\boldsymbol{q}n}+\frac{c_{\boldsymbol{q}n}}{l_{B}}\left(\hat{\gamma}_{\boldsymbol{q}n}+\hat{\gamma}_{\boldsymbol{q}n}^{\dagger}\right)\right].}
It is useful to rewrite it in terms of the canonically conjugate variables as follows:
\eqn{\label{HVU}
\frac{\hat{H}_{VU}}{\hbar\omega_{c}}=\frac{\hat{X}^{2}+\hat{P}^{2}}{2}+\sum_{\boldsymbol{q}n}\epsilon_{\boldsymbol{q}n}\frac{\hat{X}_{\boldsymbol{q}n}^{2}+\hat{P}_{\boldsymbol{q}n}^{2}}{2}-\hat{\boldsymbol{\pi}}\cdot\hat{\boldsymbol{\Xi}}_{b}+\frac{1}{2}:\hat{\boldsymbol{\Xi}}_{b}^{2}:+\sum_{\boldsymbol{q}n}\xi_{\boldsymbol{q}n}\hat{X}_{\boldsymbol{q}n},
}
where we denote the conjugate variables associated with the cyclotron motion  and the polaritons by
\eqn{
\hat{\boldsymbol{\pi}}=\left(\frac{\hat{a}+\hat{a}^\dagger}{\sqrt{2}},\frac{i(\hat{a}-\hat{a}^{\dagger})}{\sqrt{2}}\right)^{{\rm T}}=\left(\hat{X},\hat{P}\right)^{{\rm T}},\;\;
\left(\frac{\hat{\gamma}_{\boldsymbol{q}n}+\hat{\gamma}_{\boldsymbol{q}n}^{\dagger}}{\sqrt{2}},\frac{i(\hat{\gamma}_{\boldsymbol{q}n}-\hat{\gamma}_{\boldsymbol{q}n}^{\dagger})}{\sqrt{2}}\right)^{{\rm T}}=\left(\hat{X}_{\boldsymbol{q}n},\hat{P}_{\boldsymbol{q}n}\right)^{{\rm T}},
}
and introduce
\eqn{
\hat{\boldsymbol{\Xi}}_{b}=\sum_{\boldsymbol{q}n}\frac{l_{B}\boldsymbol{q}}{2}\left(\hat{X}_{\boldsymbol{q}n}^{2}+\hat{P}_{\boldsymbol{q}n}^{2}\right),\;\;\epsilon_{\boldsymbol{q}n}=\frac{\omega_{\boldsymbol{q}n}}{\omega_{c}}+\frac{l_{B}^{2}q^{2}}{2},\;\;\xi_{\boldsymbol{q}n}=\sqrt{\frac{2\omega_{\boldsymbol{q}n}}{\omega_{c}^{3}}}\frac{g_{\boldsymbol{q}n}}{ql_{B}},
}
which correspond to the dimensionless total polariton momentum, the normalized effective polariton energies, and the dimensionless coupling strengths, respectively.
The mean-field energy $\overline{E}$ is given by taking the expectation value of Eq.~\eqref{HVU} with respect to the product of coherent states for these conjugate variables. The mean-field ground state is then determined by finding the zero of the derivatives of $\overline{E}$, leading to
\eqn{
\overline{\boldsymbol{\pi}}=\overline{\boldsymbol{\Xi}}_{b}=\sum_{\boldsymbol{q}n}\frac{l_{B}\boldsymbol{q}}{2}\left(\overline{X}_{\boldsymbol{q}n}^{2}+\overline{P}_{\boldsymbol{q}n}^{2}\right),\;\;\overline{X}_{\boldsymbol{q}n}=\frac{-\xi_{\boldsymbol{q}n}}{\epsilon_{\boldsymbol{q}n}},\;\;\overline{P}_{\boldsymbol{q}n}=0.
}

\begin{figure}[t]
\includegraphics[width=160mm]{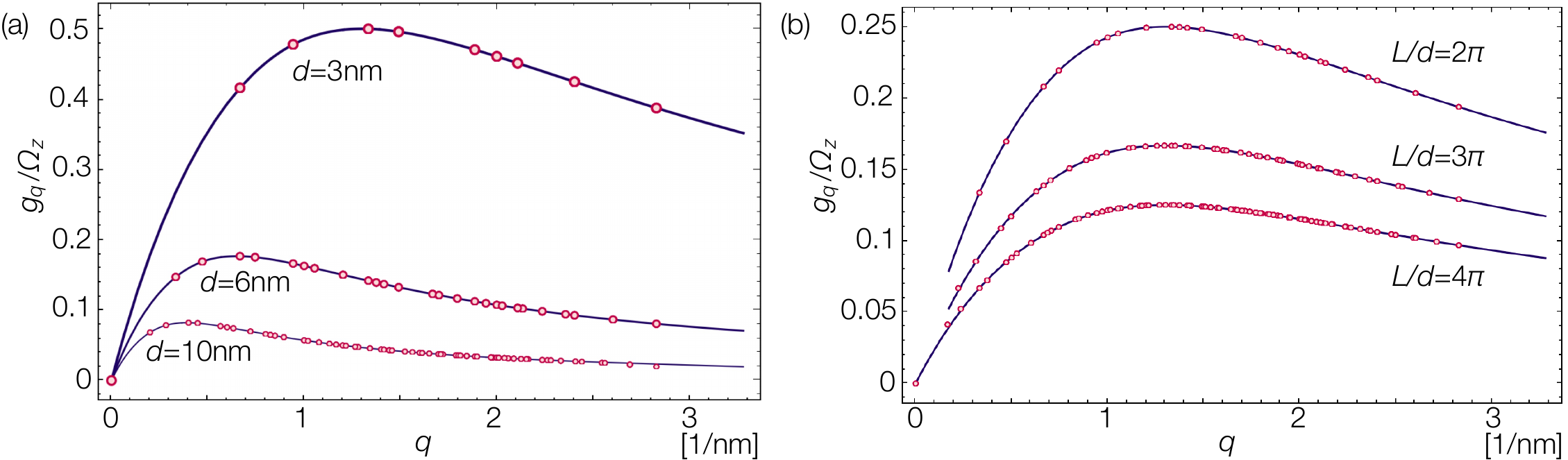} 
\caption{\label{fig_s3}
Coupling strengths $g_q$ at (a) different cavity thicknesses $d$ with $L/d=\pi$ and (b) different aspect ratios $L/d$ with $d=3{{\rm\, nm}}$. The values corresponding to each of the discrete in-plane momenta $\boldsymbol q$ of hyperbolic modes are indicated by the filled red circles. The coupling strengths indicated in panel (a) and (b) correspond to Fig.~3 in the main text and Fig.~\ref{fig_s2}, respectively. The results are plotted for the $\textit{h}$-BN parameters and the in-plane momentum cutoff $\Lambda=2{\,{\rm nm}^{-1}}$.}
\end{figure}

To obtain the low-energy excitation spectrum, we perform the projection $\hat{P}$ onto the quadratic fluctuations on top of this mean-field ground state. Specifically, we describe the low-energy excitations in terms of the effective quadratic Hamiltonian given by
\eqn{
\hat{H}_{{\rm eff}}=\hat{P}\hat{D}^{\dagger}\hat{H}_{VU}\hat{D}\hat{P},
}
where $\hat{D}$ induces the mean-field displacements
\eqn{
\hat{D}^{\dagger}\hat{\boldsymbol{\pi}}\hat{D}=\hat{\boldsymbol{\pi}}+\overline{\boldsymbol{\pi}},\;\;\hat{D}^{\dagger}\hat{X}_{\boldsymbol{q}n}\hat{D}=\hat{X}_{\boldsymbol{q}n}+\overline{X}_{\boldsymbol{q}n},\;\;\hat{D}^{\dagger}\hat{P}_{\boldsymbol{q}n}\hat{D}=\hat{P}_{\boldsymbol{q}n}.
}
The result is
\eqn{
\frac{\hat{H}_{{\rm eff}}}{\hbar\omega_{c}}&=&\frac{\hat{X}^{2}+\hat{P}^{2}}{2}+\sum_{\boldsymbol{q}n}\tilde{\epsilon}_{\boldsymbol{q}n}\frac{\hat{X}_{\boldsymbol{q}n}^{2}+\hat{P}_{\boldsymbol{q}n}^{2}}{2}+\frac{1}{2}\left[\sum_{\boldsymbol{q}n}\tilde{\xi}_{\boldsymbol{q}n}\boldsymbol{e}_{\boldsymbol{q}}\hat{X}_{\boldsymbol{q}n}\right]^{2}-\hat{\boldsymbol{\pi}}\cdot\sum_{\boldsymbol{q}n}\tilde{\xi}_{\boldsymbol{q}n}\boldsymbol{e}_{\boldsymbol{q}}\hat{X}_{\boldsymbol{q}n}\nonumber\\
&\equiv&\frac{1}{2}\hat{\boldsymbol{\phi}}^{{\rm T}}M\hat{\boldsymbol{\phi}},
}
where we introduce the modified effective polariton energies and the effective coupling strengths by
\eqn{
\tilde{\epsilon}_{\boldsymbol{q}n}\equiv\frac{\omega_{\boldsymbol{q}n}}{\omega_{c}}+\frac{l_{B}^{2}q^{2}}{2}-\frac{1}{2}l_{B}\boldsymbol{q}\cdot\overline{\boldsymbol{\Xi}}_{b},\;\;\;\tilde{\xi}_{\boldsymbol{q}n}\equiv\frac{g_{\boldsymbol{q}n}}{\omega_{c}}\sqrt{\frac{2\omega_{\boldsymbol{q}n}}{\omega_{c}}}\frac{1}{\frac{\omega_{\boldsymbol{q}n}}{\omega_{c}}+\frac{l_{B}^{2}q^{2}}{2}},
}
respectively. In the last line, we rewrite the Hamiltonian by using the matrix $M$ and the vector of the conjugate variables
\eqn{
\hat{\boldsymbol{\phi}}=\left(\hat{X},\ldots,\hat{X}_{\boldsymbol{q}n},\ldots,\hat{P},\ldots,\hat{P}_{\boldsymbol{q}n},\ldots\right)^{{\rm T}}.
}
The low-energy excitation spectrum $\left\{ \omega_{\lambda}\right\}$  can then be obtained from the Williamson eigenvalues of $M$ as follows: 
\eqn{\label{Wil}
S^{{\rm T}}MS={\rm diag}(\left\{ \omega_{\lambda}/\omega_{c}\right\} ,\left\{ \omega_{\lambda}/\omega_{c}\right\}),
}
where $S$ is a symplectic matrix. We note that this treatment is exact in the long-wavelength limit $ql_{B}\to0$ in which Eq.~\eqref{HVU} simplifies to the quadratic Hamiltonian. 

Finally, the absorption spectrum can be obtained from
\eqn{
A(\omega)&\simeq&{\rm Re}\left[\int_{0}^{\infty}e^{i\omega t}\langle0|\hat{D}^{\dagger}\hat{a}e^{-i\hat{H}_{VU}t}\hat{a}^{\dagger}\hat{D}|0\rangle\right]
\nonumber\\
&\simeq&{\rm Re}\left[\int_{0}^{\infty}e^{i\omega t}\langle0|\left(\hat{a}+\frac{\overline{\pi}_{x}-i\overline{\pi}_{y}}{\sqrt{2}}\right)e^{-i\hat{H}_{{\rm eff}}t}\left(\hat{a}^{\dagger}+\frac{\overline{\pi}_{x}+i\overline{\pi}_{y}}{\sqrt{2}}\right)|0\rangle\right].
}
The last line can easily be evaluated by using the diagonalized basis in Eq.~\eqref{Wil}.
 We note that the results in the main text are obtained for the discretized  in-plane wavenumbers as follows:
\eqn{\label{qdisc}
q_{i}\in\left\{-\Lambda,-\frac{N-1}{N}\Lambda,\ldots,-\frac{\Lambda}{N},0,\frac{\Lambda}{N},\ldots,\frac{N-1}{N}\Lambda,\Lambda\right\},\;\;i\in\{x,y\}.
}
Here, the integer number $N$ is related to the lateral system size $L$ via 
$\left[\frac{\Lambda L}{2\pi}\right]=N$. The results obtained for the cavity consisting of ultrathin $\textit{h}$-BN materials with different aspect ratios $L/d$ are plotted in Fig.~\ref{fig_s2}. The coupling strengths to each hyperbolic phonon polariton mode with discretized in-plane momentum are plotted for different thicknesses and aspect ratios  in Fig.~\ref{fig_s3}.  When the lateral size $L$ is increased, the coupling strength $g_q$ to each mode decreases uniformly as $g_q\propto L^{-1}$ while there appear a larger number of cavity modes. Consequently, a key feature of the ultrastrong coupling regime in the spectrum, that is, the formation of the localized Landau-polariton mode remains almost the same independently of the lateral size $L$. Meanwhile, the increase of $L/d$ leads to the appearance of dense anticrossed branches originating from the hybridization with the continuum cavity modes above the lower Landau-polariton mode.

\end{document}